\documentclass[aps,preprint]{revtex4}%
\usepackage{amsfonts}
\usepackage{caption}   
\usepackage{amsmath}
\usepackage{placeins}
\usepackage{float}
\usepackage{amssymb}
\usepackage{ulem}
\usepackage{orcidlink}     
\usepackage{graphicx}%
\usepackage{braket}
\providecommand{\U}[1]{\protect\rule{.1in}{.1in}}

\begin{document}
\title{Quantum-to-Quantum Transfer Learning Exhibits Improved Robustness in Low-Data Regimes}
\author{Li-An Lo}
\affiliation{Department of Engineering Science, National Cheng Kung University, Tainan 701401, Taiwan}
\author{Li-Yi Hsu}
\affiliation{Department of Physics, Chung Yuan Christian University, Chungli 32081, Taiwan and}
\affiliation{Physics Division, National Center for Theoretical Sciences, Taipei 106319, Taiwan}
\author{Hsien-Yi Hsieh\,\orcidlink{0000-0001-5227-8248}}
\affiliation{Institute of Photonics Technologies, National Tsing Hua University, Hsinchu 30013, Taiwan}
\pacs{PACS number}

\begin{abstract}
Transfer learning under limited data is a challenging setting, where models must adapt to new tasks with minimal supervision. Prior work has primarily focused on improving absolute accuracy in transfer learning. However, empirical evidence comparing quantum and classical models in realistic transfer learning settings remains limited, especially in low-data regimes. In this work, we systematically study the robustness of quantum models under reduced training data. We evaluate multiple quantum and classical architectures across diverse transfer tasks and retraining configurations, and quantify robustness using accuracy degradation and relative performance retention (RPR). Our results show that, although classical models often achieve higher peak performance, they exhibit significantly larger degradation when training data is limited. In contrast, quantum models maintain more stable performance across data regimes, indicating improved robustness and data efficiency. These findings provide empirical evidence that quantum models can offer improved robustness in low-resource transfer learning scenarios.
\end{abstract}


\maketitle

\section{Introduction}

Transfer learning (TL) is a machine learning technique that reuses a
pre-trained model for a source dataset to develop a learning model for a
target dataset, eliminating the training process from scratch \cite{survey,s1}%
. Specifically, transfer learning benefits the restricted target datasets or
medical images, where overfitting occurs in training a learning model from
scratch on these small target datasets. Transfer learning is to pre-train the
learning model on a large-scale source dataset, which is different but related
to the smaller target datasets. Next, this pre-trained model can be reused and
modified to build the learning on the target dataset. Previous studies show
that transfer learning enables training a learning model on significantly
small datasets without overfitting \cite{few,f1,f2,f3}. Transfer learning has applications in anomaly detection \cite{UTL1}, Globality-Locality Preserving Projections
\cite{UTL2}, relative distance comparisons \cite{UTL3}, convolutional sparse
coding \cite{UTL4}, target detection \cite{UTL5}, adversarial training
\cite{UTL6}, and fault diagnosis \cite{UTL7}. 

Thanks to advances in science and technology, deep neural networks can be
implemented as learning models on images in modern machine learning.
Literature reveals two consensuses on the transferability of the deep neural
networks \cite{few,f1,f2,f3}. Firstly, the transferability decreases as the
domain gap between source and target knowledge increases. Secondly, it is
observed that the earlier/lower layers close to the input learn general
knowledge while the later/higher layers close to the output learn specific
knowledge. In this case, the early layers of the pre-trained deep neural
networks\ are unchanged or frozen, while fine-tuning the later layers is
performed to learn the specific knowledge of the target dataset. In the
following, this traditional transfer learning is also called
classical-to-classical transfer learning (CtCTL) since the source and target
datasets each are trained using classical convolutional neural networks(CCNN).

With the rise of quantum information science, it is revealed that quantum
phenomena such as superposition and entanglement in a large Hilbert space
allow quantum computation to speed up over the limitations of classical
computation \cite{1,2,3}. In the noisy intermediate-scale quantum (NISQ) era,
the variational quantum algorithms (VQA) within the hybrid quantum-classical
scenario is a popular and appealing approach in quantum machine learning. To
train a quantum learning model using a VQA, a parametrized quantum circuit
(PQC) is designed, and classical optimizers are utilized to train the
parameters in the learning process. Notably, VQAs can suffer from barren
plateaus, where the gradients in the parameter landscape exponentially vanish
with the number of qubits in PQC \cite{bp,bp1,bp2}. On the other hand, the
number of qubits in quantum convolutional neural networks (QCNNs)
\cite{qcnn,qcnn1} exponentially decreases with the depth of PQCs, avoiding the
occurrences of the barren plateaus \cite{bp31,bp3,bp4}. That is, the
hierarchical structure of QCNNs prevents them from the barren plateau
phenomenon, representing a trainable architecture in quantum machine learning.
Various QCNNs have their applications in data classification \cite{good},
image classification \cite{qcnn10,qcnn11,qcnn12,qcnn4}, object detection
\cite{qcnn6,qcnn9}. Specifically, we employ two-qubit ansatz to construct the
necessary PQCs of QCNNs in the unsupervised learning tasks, yielding
classification with great accuracy using quantum circuits with shallow depth
\cite{good,ggood,cq}.

There are several types of hybrid transfer learning, training source or target
datasets using classical or models \cite{qo}. For example, in the
classical-to-quantum transfer learning, the source dataset is pre-trained
using CCNN, and then the target dataset is trained using QCNN by utilizing the
earlier layers of the pre-trained CNN for feature extraction
\cite{cq,cq1,cq2,cq3,cq4,cq5,cq6}. There are several methods for the
quantum-to-classical transfer learning (QtCTL). In \cite{qo,qc1}, the proposed
QtCTL consists of using a pre-trained QCNN as a feature extractor, and a
classical neural network used in post-processing of its outputs from the PQC.
In an alternative QtCTL, the first convolutional layer of the CNN is replaced
by a quantum circuit, where the classical weights can be easily derived from
the quantum model and then transferred to a classical machine \cite{qc}.

In this work, we investigate quantum-to-quantum transfer learning (QtQTL) as the quantum counterpart of classical-to-classical transfer learning (CtCTL) for supervised classification tasks. Specifically, a QCNN is first pretrained on a source dataset, after which its early layers are frozen while the remaining layers are retrained on a target dataset.

We systematically evaluate quantum and classical transfer learning across multiple tasks and retraining configurations, with particular emphasis on performance under limited target training data. Using relative performance retention (RPR) and accuracy degradation, we characterize the robustness of quantum models against data reduction and compare their behavior with classical baselines. Furthermore, we employ kernel-target alignment (KTA) to examine how representations learned from source tasks align with target labels, providing additional insight into the observed differences in transfer-learning performance.

The remainder of this paper is organized as follows. Section II reviews the necessary preliminaries, including quantum data encoding, parametrized quantum circuits, and QCNN architectures. Section III introduces the quantum and classical transfer-learning frameworks and describes the experimental settings. Section IV presents a systematic analysis of transfer-learning performance and robustness across all tasks, including positive-transfer scenarios, under different target training-data conditions. Section V investigates the representation-level characteristics of the pretrained models using kernel-target alignment (KTA), discusses their possible relationship to the observed robustness, and concludes the paper.
\section{Preliminaries}

\subsection{AQCNN}

\begin{figure}[t!]
\captionsetup{font=small,justification=raggedright,singlelinecheck=false}
	\begin{center}
	\includegraphics[width=.95\linewidth]{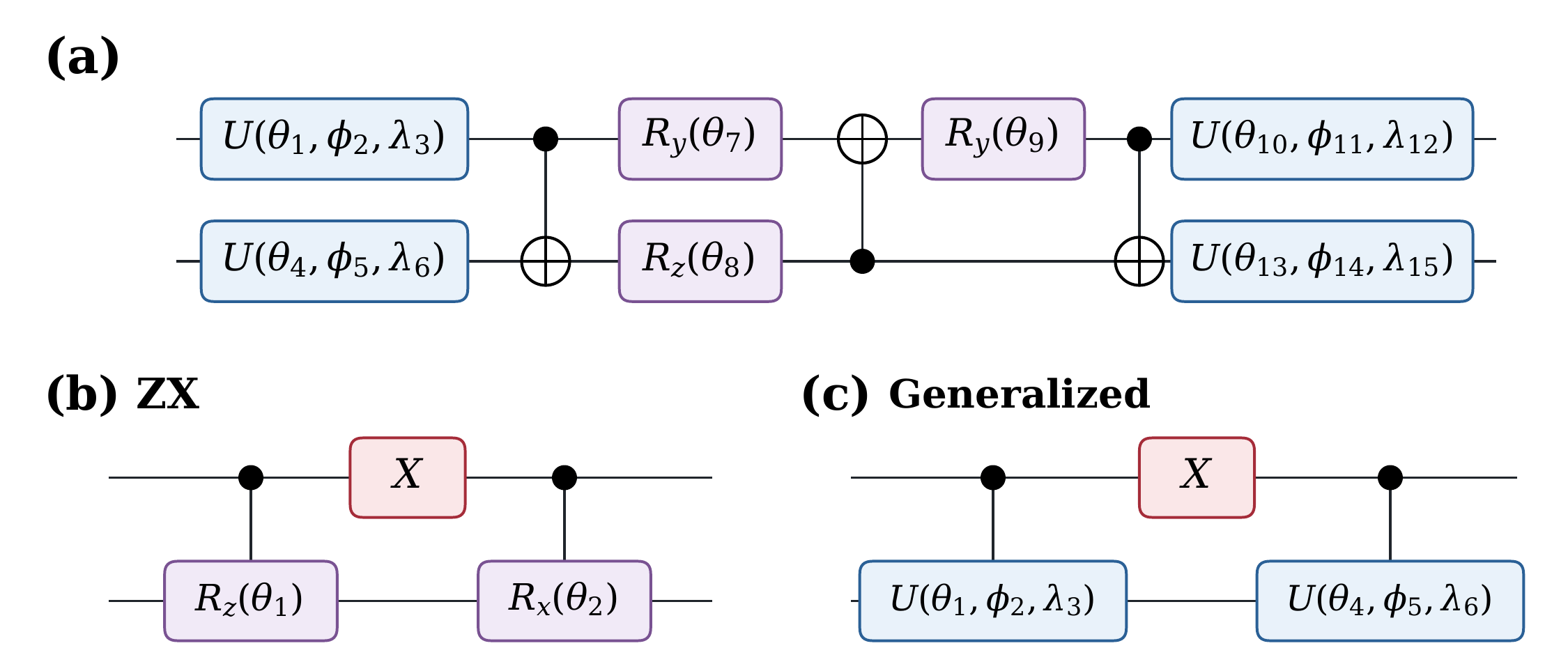}
	\end{center}
	\vskip - 1mm
    \caption{
    Ansatz circuits used in the QCNN architecture. 
    (a) Two-qubit convolutional ansatz representing a parameterization of an arbitrary $SU(4)$ unitary. 
    (b) ZX pooling ansatz. 
    (c) Generalized pooling ansatz. 
    In the pooling layers, the control qubits are discarded after the controlled-unitary operations, while the target qubits are passed to the next layer.
    }
    \label{fig:Ansatz}
\end{figure}

The architecture of a classical convolutional neural network (CNN) typically
consists of the convolutional layers and the pooling layers. A convolutional
layer is designed to extract features from input data, while a pooling layer
added after a convolutional layer is to reduce the dimensionality of the data
without harming essential information. A QCNN is a quantum-analog of classical
convolutional neural networks, also comprising convolutional and pooling
layers \cite{qcnn,qcnn1}. In this work, we focus on an ansatz-based QCNN (AQCNN), where parameterized quantum circuits (ansatzes) are used as building blocks of convolutional and pooling layers. To design the convolutional and pooling layers
in QCNNs, the two-qubit ansatzes are exploited as building blocks of these
layers. Among variant ansatzes in AQCNNs \cite{good}, the two-qubit circuit in
Figure~\ref{fig:Ansatz} (a) is quite useful for diverse learning tasks \cite{good,a2,a3} ,
representing the parameterization of an arbitrary $SU(4)$ gate \cite{a1}. In
this work, we employ this ansatz as the building block of the convolutional layers.
On the other hand, we employ the ZX pooling and the generalized pooling
ansatzes as the building blocks of pooling layers \cite{cq,good}. We make the following remarks. These two ansatzes in Figure~\ref{fig:Ansatz} (b) and (c) each comprise
controlled-unitary operation. After pooling, the controlled qubits are discarded, while the target qubits are passed to the next convolutional layer. That is, half the qubits are traced out after the
evolution of the convolutional- pooling pair, and the other survived half
enter into the next pair. In addition, the generalized pooling ansatz can
implement the parameterization of an arbitrary $SU(2)$ gate, meaning the
generalized pooling ansatz can be trained to learn ZX one \cite{cq}. In
addition, AQCNNs without the pooling layers are studied in the classification
task \cite{good} and transfer learning \cite{cq}. It is because the ansatz in
Figure~\ref{fig:Ansatz} (a) can conduct an arbitrary $SU(4)$ gate \cite{good}. Here we also
investigate the no-pooling AQCNNs. In this case, the qubits that should be
served as the controlled qubits in the pooling layers are also discarded
after the convolutional layers. Finally, the parameters of all ansatzes
employed in the convolutional layers or the pooling layer of a PQC are set the same.

\subsection{ Angle Encoding}

In this work, the classical data is embedded into quantum states using the angle
encoding method. To this end, each component $x_i$ of an $N$-dimensional data
vector $\vec{x} = [x_1, x_2, \ldots, x_N]$ is mapped to a rotation angle in a
quantum circuit, which is given by
\begin{equation}
\ket{\Phi(x)} = \frac{1}{\sqrt{N}} \bigotimes_{i=1}^{N}
R_y(x_i)\ket{0},
\end{equation}
where $R_y(x_i)$ denotes a rotation around the $y$-axis by an angle $x_i$.

\subsection{Source Data training}

The $N$-qubit ($N = 8, 16$) PQCs of the proposed AQCNN consist of initial state preparation, quantum data encoding, convolutional filters, and pooling operations. Specifically, the $i$-th qubit is initially prepared in the state $\ket{0} $, followed by a unitary rotation $R_y(x_i)$ in the data encoding stage. The resulting encoded state $\ket{\Phi(x)}$ is then processed through three variational layers of the AQCNN, each consisting of a convolutional layer followed by a pooling layer. There are $3$ and $4$ convolution--pooling layers in the $8$-qubit and $16$-qubit PQCs, respectively.  

Here we describe the $8$-qubit PQC in detail. In the first variational layer, the convolutional layer contains two sublayers. In the first sublayer, qubit pairs $(2i+1,\, 2(i+1))$ for $i = 1, \ldots, N-1$ are fed into the same ansatz. In the second sublayer, qubit pairs $(2(j+1),\, 2j+3)$ for $j = 1, \ldots, N-2$, together with the pair $(1,\, N)$, are processed by the same ansatz. After the convolutional operations, the qubit pairs $(2i+1,\, 2(i+1))$ for $i = 1, \ldots, N-1$ enter the same pooling layer, where qubits $2i+1$ and $2(i+1)$ serve as control and target qubits, respectively. The target qubits are passed to the next variational layer as input qubits, while the control qubits are discarded. The structures of the second and third variational layers are similar to that of the first layer. 

The AQCNN is trained within the framework of a variational quantum algorithm (VQA), where the parameters of the underlying parameterized quantum circuit (PQC) are optimized. After the evolution through the three variational layers, measurements are performed on the first $N-1$ qubits, while the final qubit serves as a latent representation for downstream classification. Within the hybrid quantum-classical optimization framework, classical optimizers are used to update the PQC parameters in order to minimize a cost function. In this work, the Adam optimizer is employed to minimize a binary cross-entropy (BCE) loss defined over these measured qubits.

Specifically, denote the true label as $y$ and let $p_i = \Pr(\ket{x_i} = \ket{0})$ be the probability of measuring the $i$-th qubit in the $\ket{0}$ state for $i = 1, \dots, N-1$. The BCE loss is defined as
\begin{equation}
\mathcal{L}_{\mathrm{BCE}} = -\frac{1}{T} \sum_{i=1}^{T} \left[ y \log_2 p_i + (1 - y)\log_2(1 - p_i) \right],
\end{equation}
where $T = N-1$. This objective encourages the first $N-1$ qubits to collapse toward $\ket{0}$, effectively compressing the relevant information into the final qubit.

\begin{figure}[t!]
\captionsetup{font=small,justification=raggedright,singlelinecheck=false}
\begin{center}
\includegraphics[width=.95\linewidth]{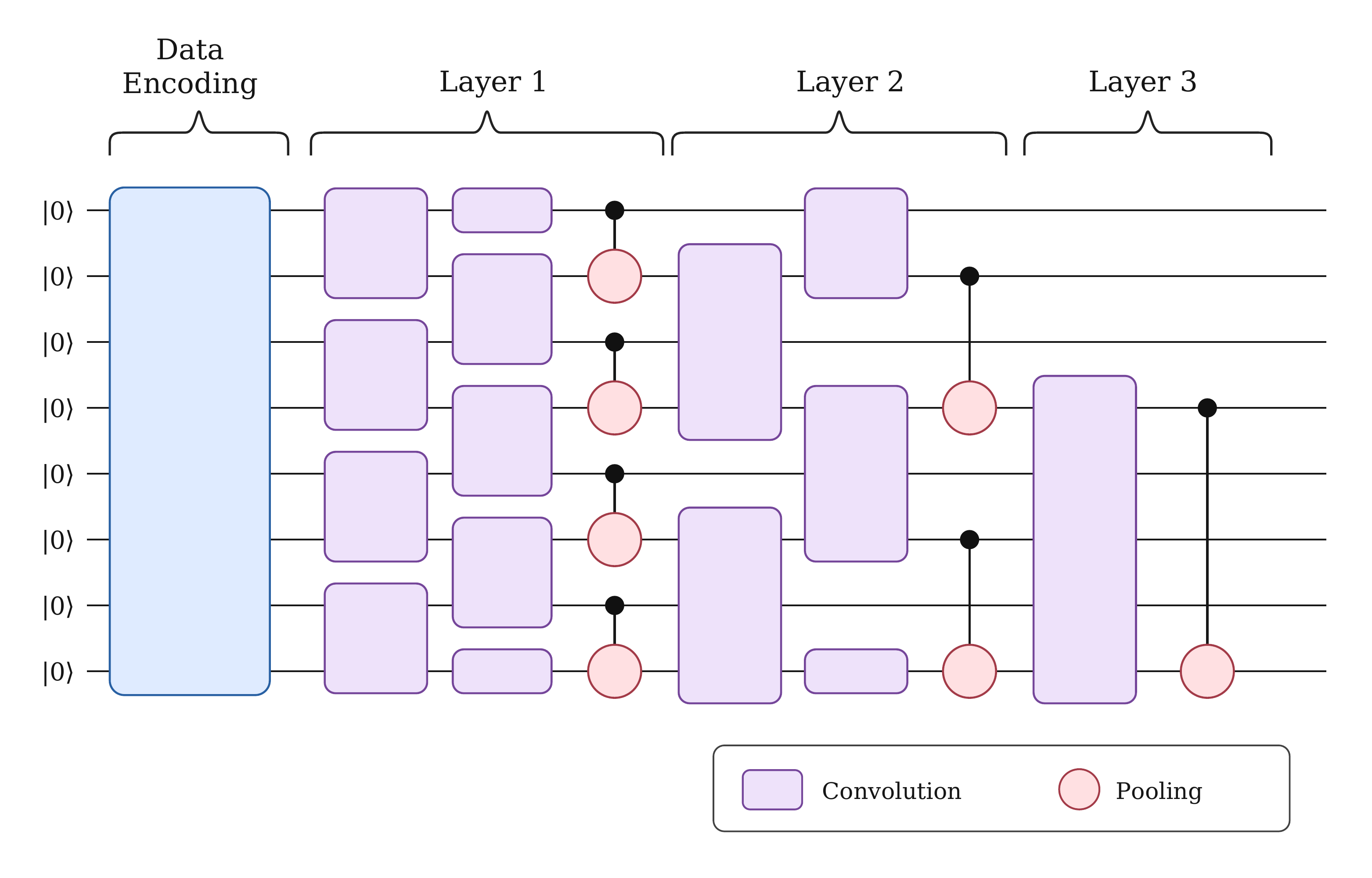}
\end{center}
\vskip - 1mm
\caption{
Schematic of the AQCNN architecture for $N = 8$ qubits. 
The circuit consists of three variational layers, each comprising a convolutional layer followed by a pooling layer. 
Input data are embedded via angle encoding using single-qubit rotations, and processed through parameterized two-qubit ansatz circuits. In each pooling layer, controlled-unitary operations are applied, after which the control qubits are traced out while the target qubits are propagated to the next layer, leading to a hierarchical reduction in system size.
}
\label{fig:architecture}
\end{figure}

After training, the final qubit is interpreted as a low-dimensional representation and mapped onto the Bloch sphere via its expectation values. Since the learned representation is not necessarily aligned with the measurement axis, a classical support vector machine (SVM)\cite{SVM} is applied to determine the optimal separating hyperplane. The resulting normal vector is then translated into a single-qubit rotation gate, aligning the decision boundary with the computational basis, such that classification can be performed by measuring the final qubit. This design separates representation learning from classification, improving flexibility in decision boundary alignment. Figure~\ref{fig:architecture} illustrates the schematic of the PQC for the case $N = 8$. Finally, the architecture of the $16$-qubit PQC is analogous to that of the $8$-qubit PQC.

\subsection{Target Data training}

After training on the source dataset, the trained PQC is transferred to the target task for further optimization. In the $8$-qubit PQCs, the parameters in the early one or two variational layers are frozen, while the other later layers are updated. Similarly, in the $16$-qubit PQCs, the parameters in the early one, two or three variational layers are frozen, while the other later layers are jointly updated. 

Given the source dataset consisting of 12{,}000 samples, we consider two target dataset sizes—a large dataset with 12{,}000 samples and a small dataset with 40 samples—to evaluate robustness under data scarcity.

\subsection{Positive transfer}

We define positive transfer separately for each task-model pair. For a given task \(t\) and model \(j\), positive transfer occurs when the best accuracy achieved with partial retraining (\(m<n\)) exceeds that obtained by training from scratch (\(m=n\)):

\[
\Delta_{t,j}^{(n)}
=
\max_{0\leq m<n}
\mathrm{Acc}_{t,j}^{\mathrm{large}}(m)
-
\mathrm{Acc}_{t,j}^{\mathrm{large}}(n)
>0.
\]

Here, \(\mathrm{Acc}^{\mathrm{large}}\) denotes accuracy evaluated using the larger target training dataset. This model-specific criterion is applied consistently across evaluations, with positive-transfer subsets determined independently of small-data performance.

\section{QUANTUM AND CLASSICAL TRANSFER-LEARNING FRAMEWORKS}

We compare the accuracy of QtQTL and CtCTL under different model configurations. 
Three key factors are considered: (i) the number of convolution--pooling layers, 
(ii) the number of parameters, and (iii) the number of retrained layers together 
with the data size used for training on the target dataset.

For QtQTL, we employ ansatz-based PQCs with $n$ convolution--pooling layers in the $2^{n}$-qubit PQCs ($n=3,4$), and let $m$ denote the number of retrained layers. The transfer learning procedure is defined as follows. 
Given an $n$-layer model, the first $m$ layers are fixed while the remaining $n-m$ layers are retrained ($0 \leq m \leq n$, $n=3,4$). The case $m=n$ corresponds to training from scratch on the target dataset.
In this work, we investigate three ansatz-based QCNN variants, including generalized pooling (QCNN-G), ZX pooling (QCNN-Z), and no pooling (QCNN-N). 
All these QCNN variants involve fewer than 100 parameters.

For CtCTL, we consider two types of classical CNNs (CCNNs) with different model capacities: CCNN-A, which has a comparable number 
of parameters to the QCNN models, and CCNN-B, which has a larger number of parameters 
and serves as a strong classical baseline. 
The number of parameters for all models is summarized in Table~\ref{tab:params}. Therein, CCNN-A is designed to match the parameter scale of QCNN variants, enabling a controlled comparison. 
In addition, CCNN-B is included as a stronger baseline to avoid underestimating classical performance. 

To examine the effect of data size, we consider two target datasets: a large dataset 
with 12{,}000 samples and a small dataset with 40 samples. 
The source dataset size is fixed at 12{,}000 samples, as described in Section~2. 

\begin{table}[H]
\centering
\small
\caption{Number of parameters for different models.}
\label{tab:params}
\begin{tabular}{c|c|c}
\hline
 & $n=3$ & $n=4$ \\
\hline
QCNN-N & 45 & 60 \\
QCNN-Z & 51 & 68 \\
QCNN-G & 63 & 84 \\
CCNN-A & 49 & 185 \\
CCNN-B & 11901 & 22001 \\
\hline
\end{tabular}
\end{table}
We study transfer learning tasks using the MNIST and Fashion-MNIST datasets as benchmarks.
The original $28 \times 28$ images are reduced to 8 and 16 dimensions for the $n=3$ 
and $n=4$ cases, respectively, using the method proposed in \cite{ggood}. 
In the first six transfer learning tasks, binary classification is performed by 
selecting pairs of digit classes (e.g., $\{"0","8"\}$, $\{"5","7"\}$, and $\{"1","2"\}$). 
In the remaining tasks, models are first trained for multi-class classification and 
then adapted to binary classification.

The best classification accuracies for each transfer-learning task are summarized in the tables in Appendix A. A more comprehensive analysis based on the full set of numerical results is presented in the following section. All numerical simulations were implemented using PennyLane and executed on CPUs. Classification accuracy was evaluated on a test set of 400 samples.

We also performed a limited hardware experiment using the Quantinuum H2-1 quantum processor \cite{quantinuum_h1}. Specifically, we considered a transfer-learning task from digits ``1'' and ``2'' to digits ``0'' and ``8''. For three test images of digit ``0'', we estimated the probability of predicting class ``0'' using 1,000 measurement shots per image. The simulator predictions were $89.90\%$, $82.48\%$, and $82.85\%$, while the corresponding hardware results were $90.10\%$, $81.10\%$, and $81.90\%$, respectively. The mean absolute difference between simulator and hardware predictions was approximately $0.84$ percentage points. Although limited in scope, these results show preliminary agreement between hardware and simulator predictions for the tested inputs.

Overall, the numerical results suggest that, although quantum models do not consistently outperform classical baselines in absolute classification accuracy, they can exhibit improved performance retention under limited target training data. We examine this behavior more systematically in the following section.

\section{PERFORMANCE AND ROBUSTNESS ANALYSIS}
\subsection{Robustness under Limited Data}

We evaluate model robustness under reduced training data using the absolute accuracy drop as a summary metric. The accuracy drop is defined as the difference between accuracy under large and small training data; the mean absolute accuracy drop, aggregated across all transfer tasks and retraining depths, is reported for both all tasks and positive-transfer subsets.

\begin{figure}[t!]
\captionsetup{font=small,justification=raggedright,singlelinecheck=false} 

	\begin{center}
	\includegraphics[width=.95\linewidth]{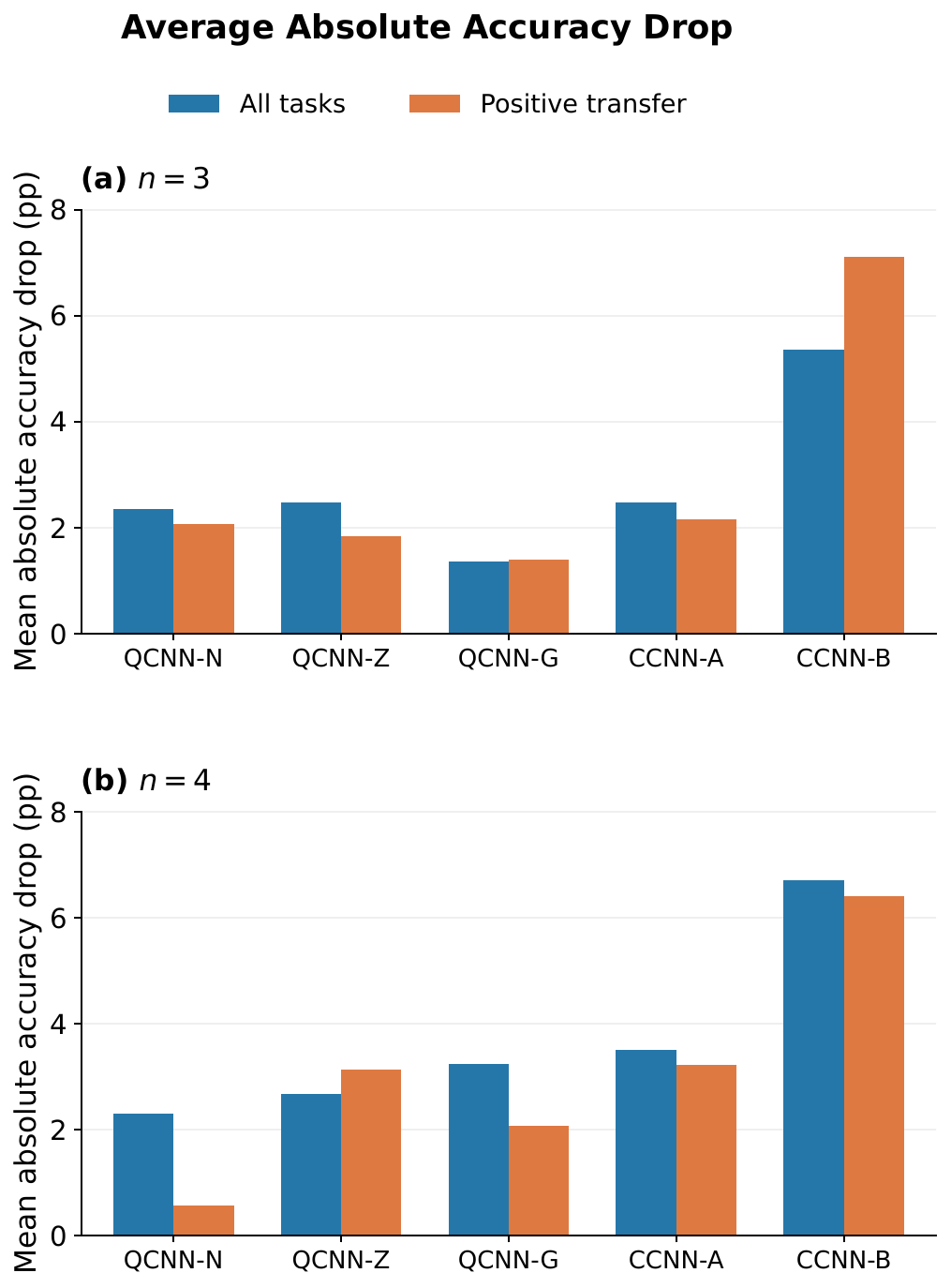}
	\end{center}
	\vskip - 1mm
	\caption{Average absolute accuracy drop across models under reduced training data. The accuracy drop is defined as the absolute difference in classification accuracy between small (40 samples) and large (12,000 samples) target training datasets, averaged over transfer-learning tasks and retraining depths \(m<n\), consistent with the subsequent RPR analysis. Results are shown for \(n=3\) (top) and \(n=4\) (bottom), considering both all tasks and model-specific positive-transfer subsets. Across both settings, quantum models (QCNN variants) generally exhibit smaller accuracy differences than classical baselines, suggesting greater robustness under limited data. The unusually low positive-transfer accuracy drop of QCNN-N at \(n=4\) should be interpreted with caution, as its positive-transfer subset contains only one task.
}
	\label{fig:abs_drop}
\end{figure}

As summarized in Figure~\ref{fig:abs_drop}, quantum models tend to exhibit smaller average accuracy degradation compared to classical baselines across both $n=3$ and $n=4$ layer settings. This overall trend is observed in the all-task analysis. When focusing specifically on positive-transfer tasks, we find that quantum models are more likely to retain stronger performance under limited data, suggesting improved robustness in these scenarios.
\subsection{Performance Retention Across Retraining Depths (${\lowercase{n}}=3$)}

The RPR is defined as the ratio between accuracy under small and large target data. Here, ``large'' and ``small'' refer to the two target dataset sizes (12,000 and 40 samples),  respectively. This ratio measures the fraction of performance preserved under reduced training data and provides a scale-invariant measure of robustness across tasks with different baseline accuracies. This also allows us to distinguish whether reduced performance degradation arises from inherently better generalization or simply lower baseline accuracy
\begin{equation}
\mathrm{RPR} = \frac{\mathrm{Accuracy}_{\text{small}}}{\mathrm{Accuracy}_{\text{large}}}
\end{equation}
An RPR value closer to 1 indicates better robustness, meaning that the model preserves a larger fraction of its original performance under reduced training data. 

\begin{figure}[t!]
\captionsetup{font=small,justification=raggedright,singlelinecheck=false} 

	\begin{center}
	\includegraphics[width=.95\linewidth]{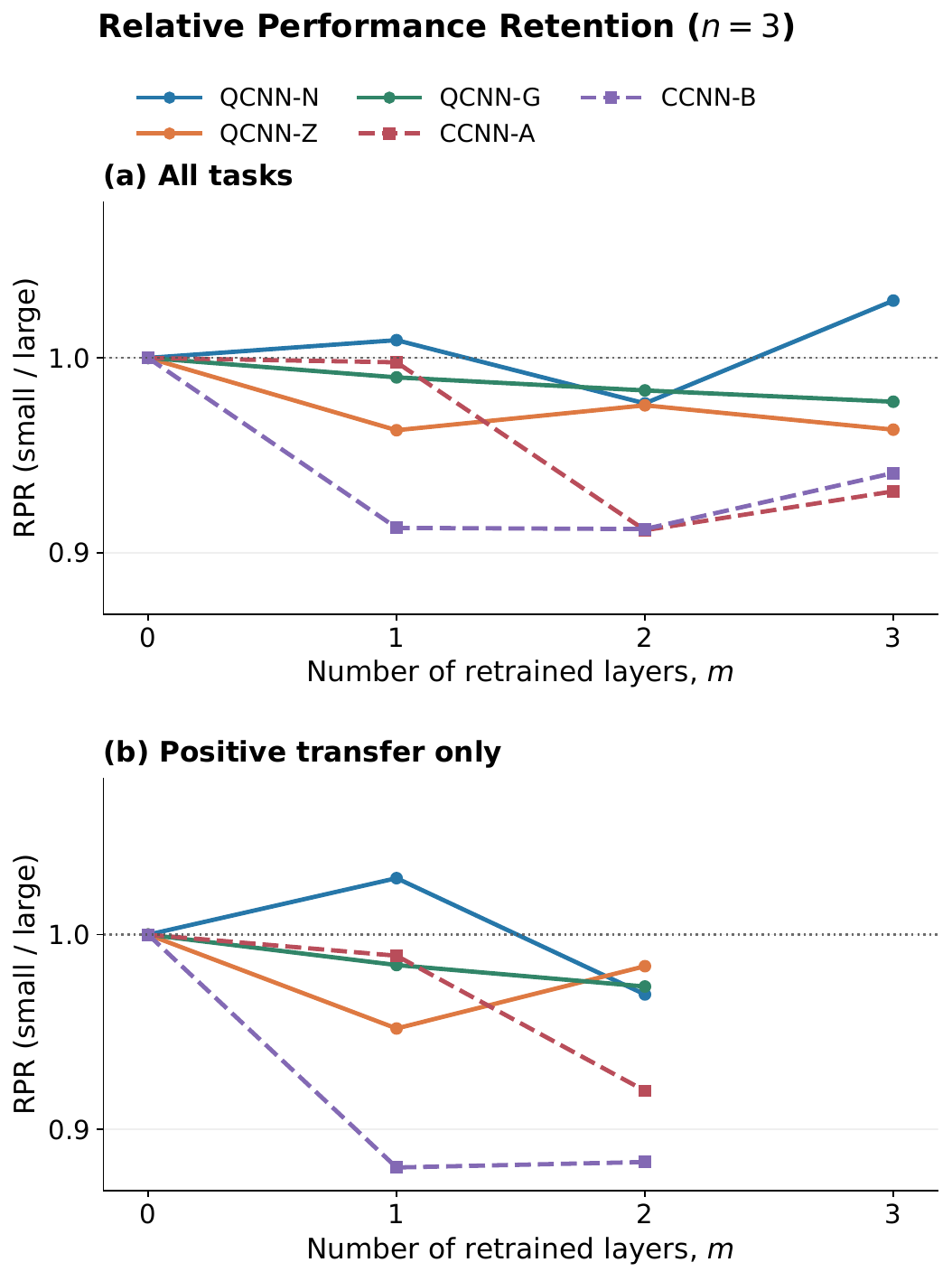}
	\end{center}
	\vskip - 1mm
	\caption{
Relative performance retention (RPR) for $n=3$, defined as the ratio between accuracy under reduced (small) and large training data. 
RPR is shown as a function of the number of retrained layers $m$ for all tasks (left) and positive-transfer tasks only (right). 
Positive-transfer tasks are defined at the task level (see text for details). Quantum models generally maintain higher retention than classical baselines, indicating reduced sensitivity to limited data.
}
	\label{fig:rpr_n3}
\end{figure}

To further understand this behavior, Figure~\ref{fig:rpr_n3} presents the relative performance retention (RPR) as a function of the number of retrained layers $m$ for the $n=3$ setting.

Across all tasks (Figure~\ref{fig:rpr_n3}, left), quantum models maintain higher RPR values than classical baselines over a wide range of retraining depths. The qualitative trends are consistent with those observed using the average absolute accuracy drop. Quantum models also preserve a larger fraction of their original performance when training data is reduced. In particular, the phenomenon where quantum models suffer a smaller drop in RPR becomes more pronounced as the number of retrained layers increases, implying that quantum models can better handle increased model flexibility without overfitting in low-data regimes. This suggests that quantum models are less sensitive to both data scarcity and the extent of fine-tuning.

In the all-tasks analysis, we explicitly retain the $m = n$ setting as a baseline, enabling direct comparison between transfer learning and training-from-scratch regimes. Even when compared to full retraining ($m = n$), quantum models still exhibit smaller performance degradation under limited data.

To examine whether robustness under limited training data contributes to transferability, we specifically isolate the results for positive-transfer tasks (Figure~\ref{fig:rpr_n3}, right). Overall, the results largely confirm this relationship, particularly as the number of retrained layers increases.
\begin{figure}[t!]
\captionsetup{font=small,justification=raggedright,singlelinecheck=false} 
	\begin{center}
	\includegraphics[width=.95\linewidth]{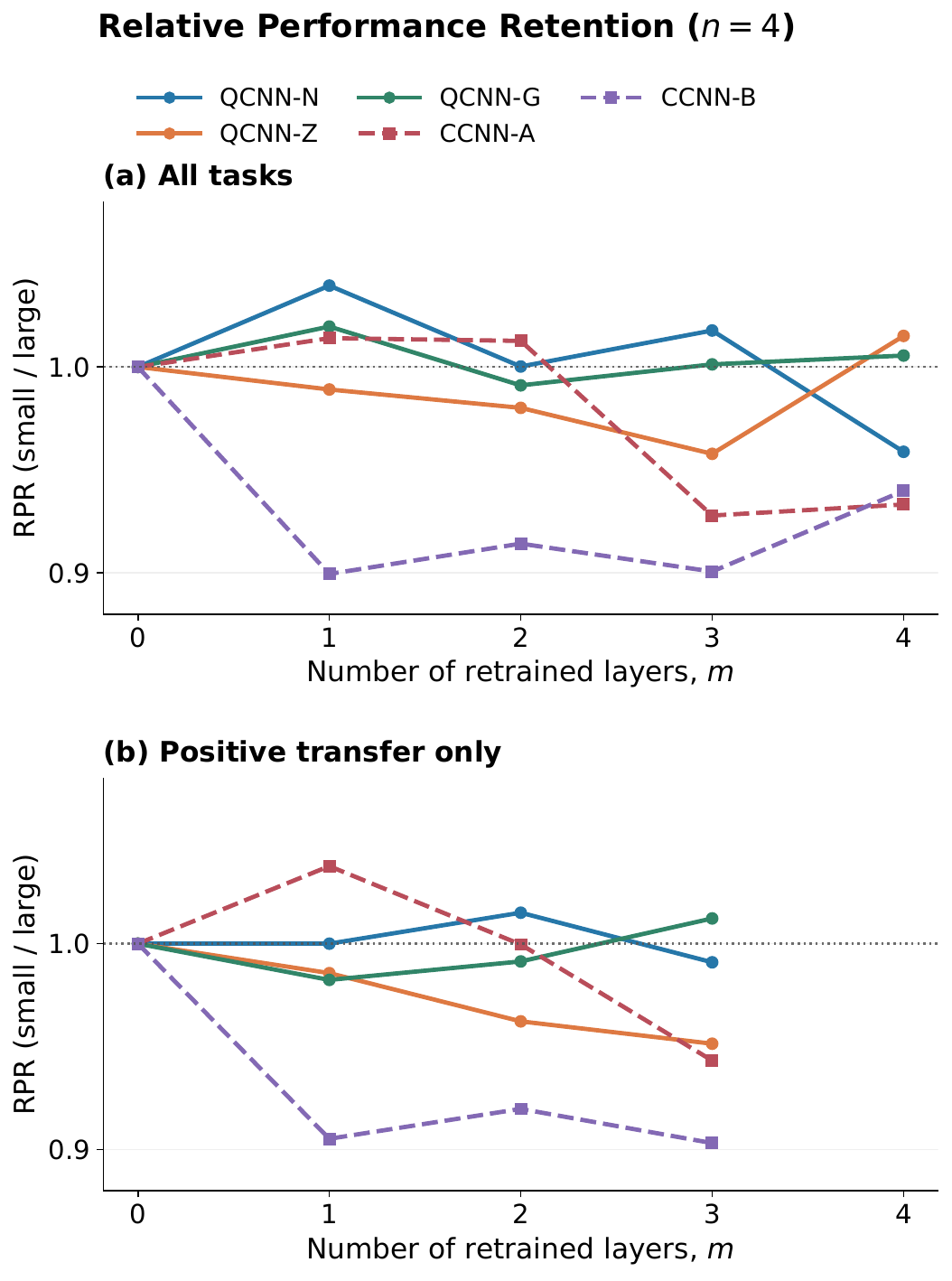}
	\end{center}
	\vskip - 1mm
	\caption{
Relative performance retention (RPR) for $n=4$, defined as the ratio between accuracy under reduced (small) and large training data. 
RPR is plotted against the number of retrained layers $m$ for all tasks (left) and positive-transfer tasks only (right). Positive-transfer tasks are defined at the task level (see text for details). The overall trend is consistent with the $n=3$ setting, where quantum models exhibit stronger performance retention under limited data. }
	\label{fig:rpr_n4}
\end{figure}
\subsection{Consistency Across Model Scales (${\lowercase{n}}=4$)}
Figure~\ref{fig:rpr_n4} shows the corresponding RPR results for the $n=4$ setting. The overall trend remains consistent with that observed for $n=3$: quantum models generally exhibit stronger performance retention under limited data.

Similarly, the analysis for \(n=4\) restricted to positive-transfer tasks also demonstrates that when more layers are retrained, the robustness of quantum models to limited data contributes to positive transfer.

\subsection{Summary of Findings}
Overall, our results reveal a consistent pattern:

\begin{itemize}
    \setlength\itemsep{0.3em}
    \renewcommand\labelitemi{--}

    \item Quantum models exhibit \textbf{smaller performance degradation} under reduced training data.
    \item This advantage is consistently observed in both the aggregated performance summary (Figure~\ref{fig:abs_drop}) and the RPR trends across retraining depths (Figures~\ref{fig:rpr_n3} and~\ref{fig:rpr_n4}).
    \item The trend holds across different model sizes ($n=3$ and $n=4$).
\end{itemize}

These findings highlight a potential advantage of quantum models in low-resource transfer learning scenarios, where maintaining performance under limited data is critical. While classical models often achieve higher peak performance, they exhibit larger degradation under reduced data. In contrast, quantum models maintain more stable performance across data regimes, suggesting improved robustness and data efficiency in transfer learning settings. 
This observation aligns with our empirical findings across both aggregate and layer-wise analyses.

\section{Representation-Level Analysis}
\subsection{Kernel-Target Alignment}
The robustness of QCNNs under limited training data may be partly attributed to their architectural inductive biases, including parameter sharing across convolutional filters and hierarchical pooling operations. These structural features may act as implicit regularization, constraining the effective hypothesis space and reducing sensitivity to limited training data.

However, architectural regularization alone may not fully account for the effectiveness of knowledge transfer between different tasks. Another potentially important factor is the compatibility between the representations learned from the source task and the classification structure of the target task. In particular, effective transfer may be facilitated when source-pretrained representations preserve features that are informative for distinguishing target classes. This motivates us to investigate the alignment between source-pretrained representations and target labels. To quantify this alignment, we employ kernel-target alignment (KTA)\cite{KTA,KTA_review}, which measures the correspondence between the similarity structure induced by learned representations and the target labels.

\begin{equation}
A(K,K_Y)
=
\frac{
\left\langle
\widetilde{K},
\widetilde{K}_Y
\right\rangle_F
}{
\|\widetilde{K}\|_F
\,
\|\widetilde{K}_Y\|_F
}
\label{eq:KTA}
\end{equation}

In Eq. (\ref{eq:KTA}), \(K\) denotes either the classical kernel \(K_C\) or the quantum kernel \(K_Q\), constructed from representations obtained after source-task training. For each transfer-learning task, a set of \(N_{\mathrm{probe}}\) labeled samples is drawn from the target dataset to evaluate the representations at different transfer boundaries, without performing target-task retraining. At \(m=n\), the representations correspond to the encoded input features, whereas for \(m<n\), they are obtained after the retained source-pretrained layers. The resulting kernel matrix \(K\in\mathbb{R}^{N_{\mathrm{probe}}\times N_{\mathrm{probe}}}\) characterizes the pairwise similarities among these probe samples, while \(K_Y=yy^{\mathrm T}\) denotes the corresponding label kernel, where \(y\in\{-1,+1\}^{N_{\mathrm{probe}}}\) contains the binary target labels. To maintain consistency with our other evaluations, we sample 400 test instances, as preliminary experiments with 200 and 1,000 samples yielded comparable results.

Before computing the alignment, both kernel matrices are centered according to \(\widetilde K=HKH\) and \(\widetilde K_Y=HK_YH\), where \(H=I-\frac{1}{N_{\mathrm{probe}}}\mathbf{1}\mathbf{1}^{\mathrm T}\) is the centering matrix and \(I\) is the identity matrix. The brackets \(\langle A,B\rangle_F=\mathrm{Tr}(A^{\mathrm T}B)\) and \(\|A\|_F=\sqrt{\langle A,A\rangle_F}\) denote the Frobenius inner product and norm, respectively. The normalized alignment therefore quantifies the correspondence between the similarity structure of the source-pretrained representations and the target-label structure. 
\begin{figure}[t!]
\captionsetup{font=small,justification=raggedright,singlelinecheck=false} 
	\begin{center}
	\includegraphics[width=.95\linewidth]{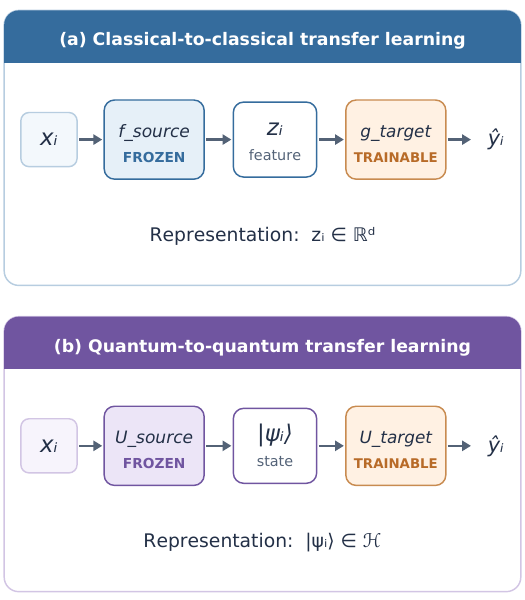}
	\end{center}
	\vskip - 1mm
	\caption{
Two transfer learning scenarios: (a) classical-to-classical and (b) quantum-to-quantum. The intermediate representations at the transfer boundary are a classical feature vector \(z_i\in\mathbb{R}^d\) and a quantum state \(|\psi_i\rangle\in\mathcal{H}\), respectively.
}
	\label{fig:Fig6}
\end{figure}

As illustrated in Fig.~\ref{fig:Fig6}, target probe samples are passed through the source-pretrained layers of the classical and quantum models to extract the corresponding hidden representations, \(z_i\) and \(|\psi_i\rangle\), respectively. These representations are then used to construct the classical and quantum kernel matrices.

For quantum models, we adopt the state fidelity as the kernel function, since the hidden representations are quantum states. The quantum kernel is defined as
\begin{equation}
K_Q(i,j)=|\langle\psi_i|\psi_j\rangle|^2,
\label{eq:quantum_kernel}
\end{equation}
where \(|\psi_i\rangle\) and \(|\psi_j\rangle\) denote the quantum representations of two target probe samples.

To facilitate comparison with the quantum kernel, we construct the classical kernel using the squared inner product of normalized feature vectors,
\begin{equation}
K_C(i,j)=
\left|\left\langle\hat z_i,\hat z_j\right\rangle\right|^2,
\qquad
\hat z_i=\frac{z_i}{\|z_i\|_2},
\label{eq:classical_kernel}
\end{equation}
where \(z_i\) denotes the classical hidden representation. This normalization gives the classical kernel the same squared-inner-product form as the quantum fidelity kernel, with both kernels taking values between zero and one.

\begin{figure}[t!]
\captionsetup{font=small,justification=raggedright,singlelinecheck=false} 

	\begin{center}
	\includegraphics[width=.95\linewidth]{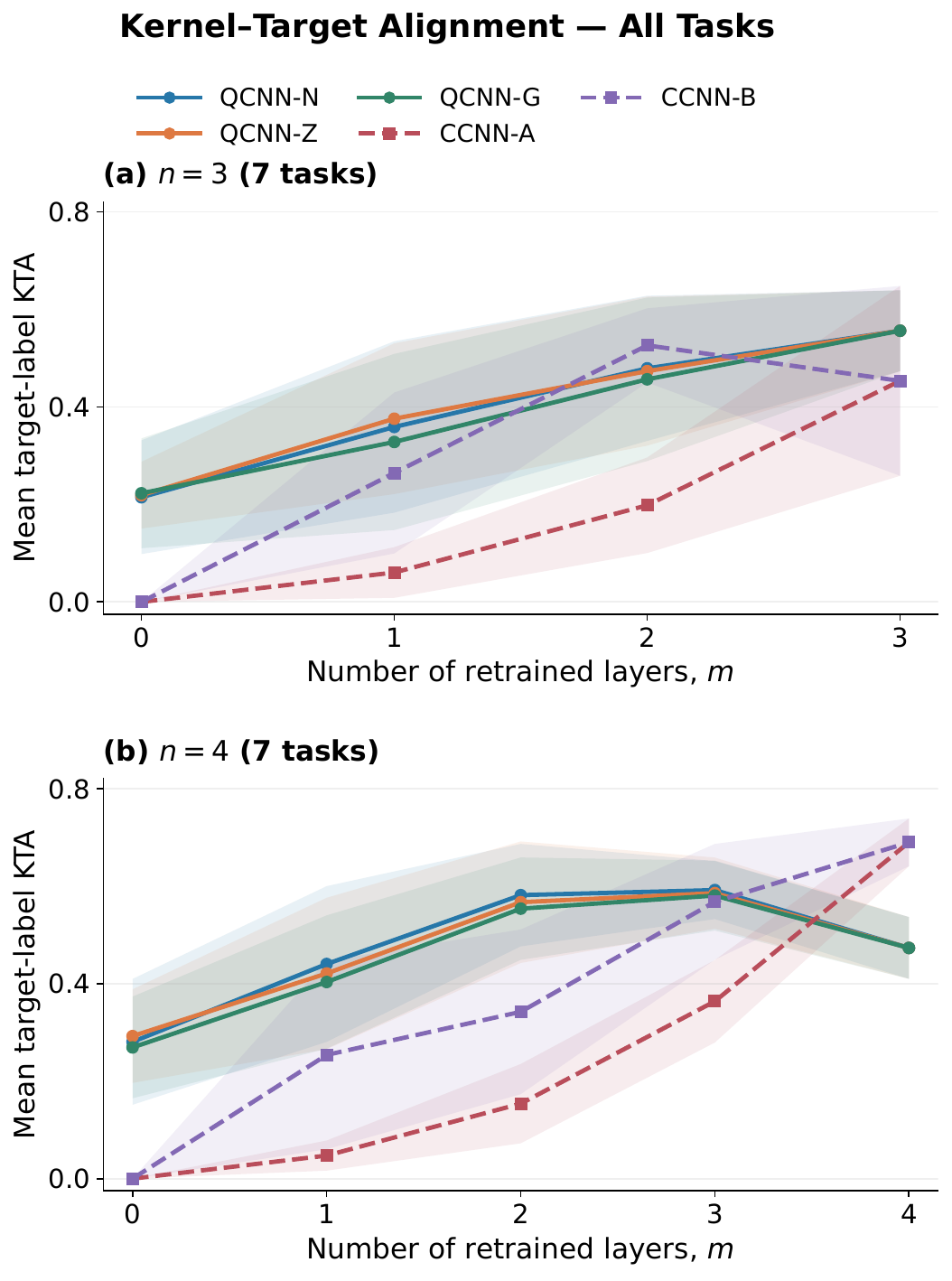}
	\end{center}
	\vskip - 1mm
	\caption{Kernel-target alignment (KTA) averaged over seven transfer-learning tasks for (a) \(n=3\) and (b) \(n=4\). The shaded regions represent ±1 standard deviation across the seven transfer-learning tasks. At \(m=n\), all layers are retrained without transfer learning, and the kernels are computed directly from the encoded input data. Consequently, the KTA values converge to a common baseline within each model family, while the quantum and classical baselines differ due to their distinct kernel functions. As \(m\) decreases, more source-pretrained layers are retained. Although the degradation of classical KTA is partly associated with architectural characteristics, such as the decreasing number of neurons in deeper layers, the results nevertheless show that classical models exhibit a more rapid decline in target-label alignment as the depth of source-pretrained layers increases.}
	\label{Fig7}
\end{figure}

To examine representation-level characteristics at different transfer boundaries, we evaluate KTA using different numbers of retained source-pretrained layers, corresponding to different retraining depths \(m\). Importantly, KTA is computed from source-pretrained representations before any target-task retraining, whereas the relative performance retention (RPR) reported earlier is evaluated after target-task retraining. These two metrics therefore characterize distinct aspects of transfer learning: representation alignment with target labels and robustness to limited target training data, respectively. The resulting KTA values are presented in Fig.~\ref{Fig7}.

\subsection{Interpretation of KTA and Transfer Robustness}

Starting from \(m=n\), where no pretrained layers are retained, the KTA values characterize the kernels constructed from the encoded input features. The differences in alignment between quantum and classical models may arise from their distinct feature encodings and kernel constructions.

As \(m\) decreases, more source-pretrained layers are retained, increasing the influence of source-task specialization and generally reducing KTA. Nevertheless, the QCNN variants exhibit a more gradual decline in KTA at several intermediate transfer depths. Since classical representations may become low-dimensional at certain depths, potentially leading to degenerate kernel matrices, we do not interpret the differences at \(m=0\) as evidence of quantum advantage. Instead, we emphasize the slower decline in KTA within \(1<m<n\), which may favor transfer learning.

The decrease in RPR with increasing \(m\) can be partly attributed to the growing number of layers that must be retrained using limited target data, increasing the estimation burden when only 40 training samples are available. Nevertheless, for positive-transfer tasks, the QCNN variants generally exhibit smaller performance degradation as the target training set is reduced. We emphasize that representation alignment and limited-data robustness capture distinct aspects of transfer learning.

\section{Conclusion}

In this work, we systematically investigated the performance and robustness of quantum-to-quantum transfer learning under limited target training data. Our results show that QCNN variants generally exhibit greater robustness to reduced training data than classical baselines, a trend that holds true even in positive-transfer settings. Furthermore, our analyses of kernel-target alignment and relative performance retention suggest a possible connection between pretrained representation geometry and limited-data transfer robustness. However, differences in model architectures and kernel constructions limit direct comparisons of representation alignment. Future research directions could involve investigating this connection more systematically to further quantify the underlying mechanisms governing the cross-task transferability of quantum models.

\clearpage
\section{Appendix}
This appendix presents the classification accuracies obtained across the transfer-learning tasks, model architectures, and retraining depths considered in this study. The tables summarize the best-performing accuracies, with the corresponding number of retrained layers \(m\) indicated in parentheses.

Bold values indicate the best performance among all models, while italic values indicate the best performance among the QCNN variants. Bold italic values highlight cases in which a quantum model outperforms the classical baselines.

All numerical simulations were implemented using PennyLane and executed on CPU resources. The computations were performed on a system equipped with an AMD Ryzen 9 7950X processor and 64 GB of RAM. Classification accuracy was evaluated using a test set of 400 samples.
\FloatBarrier
\subsubsection*{TL (I).}
Binary classification between digits ``1'' and ``2'' (source) and ``5'' and ``7'' (target).
\vspace{1mm}
\vspace{1mm}

\begin{table}[H]
\centering
\footnotesize
\setlength{\tabcolsep}{4pt}
\caption{Best accuracies for TL (I) with $n=3$.}
\label{tab:tl1_n3}
\resizebox{\columnwidth}{!}{%
\begin{tabular}{c|c|c|c|c|c}
\hline
 & QCNN-N & QCNN-Z & QCNN-G & CCNN-A & CCNN-B \\
\hline
Large data & 72.25\% (3) & \textit{81.25\%} (3) & 76.88\% (3) & 78.99\% (3) & \textbf{98.24\%} (3) \\
Small data & \textit{76.81\%} (3) & 74.25\% (3) & 76.19\% (2) & 71.21\% (3) & \textbf{86.20\%} (3) \\
\hline
\end{tabular}%
}
\end{table}

\begin{table}[H]
\centering
\footnotesize
\setlength{\tabcolsep}{4pt}
\caption{Best accuracies for TL (I) with $n=4$.}
\label{tab:tl1_n4}
\resizebox{\columnwidth}{!}{%
\begin{tabular}{c|c|c|c|c|c}
\hline
 & QCNN-N & QCNN-Z & QCNN-G & CCNN-A & CCNN-B \\
\hline
Large data & \textit{80.19\%} (4) & 69.31\% (3) & 73.63\% (4) & 77.40\% (3) & \textbf{98.79\%} (3) \\
Small data & 74.28\% (3) & 71.95\% (4) & \textit{78.62\%} (3) & 73.50\% (1) & \textbf{91.89\%} (4) \\
\hline
\end{tabular}%
}
\end{table}
\FloatBarrier

\FloatBarrier
\subsubsection*{TL (II).}
Binary classification between digits ``5'' and ``7'' (source) and ``1'' and ``2'' (target).
\vspace{1mm}
\vspace{1mm}

\begin{table}[H]
\centering
\footnotesize
\setlength{\tabcolsep}{4pt}
\caption{Best accuracies for TL (II) with $n=3$.}
\label{tab:tl2_n3}
\resizebox{\columnwidth}{!}{%
\begin{tabular}{c|c|c|c|c|c}
\hline
 & QCNN-N & QCNN-Z & QCNN-G & CCNN-A & CCNN-B \\
\hline
Large data & 96.94\% (4) & \textit{96.94\%} (3) & 97.56\% (4) & 82.78\% (3) & \textbf{98.08\%} (3) \\
Small data & 92.12\% (3) & \textit{\textbf{96.75\%}} (4) & 96.50\% (3) & 73.60\% (1) & 94.91\% (4) \\
\hline
\end{tabular}%
}
\end{table}

\begin{table}[H]
\centering
\footnotesize
\setlength{\tabcolsep}{4pt}
\caption{Best accuracies for TL (II) with $n=4$.}
\label{tab:tl2_n4}
\resizebox{\columnwidth}{!}{%
\begin{tabular}{c|c|c|c|c|c}
\hline
 & QCNN-N & QCNN-Z & QCNN-G & CCNN-A & CCNN-B \\
\hline
Large data & \textit{91.06\%} (4) & 87.38\% (3) & 82.25\% (4) & 90.40\% (4) & \textbf{98.75\%} (4) \\
Small data & 82.00\% (4) & \textit{89.12\%} (4) & 88.38\% (3) & 80.19\% (3) & \textbf{94.12\%} (4) \\
\hline
\end{tabular}%
}
\end{table}
\FloatBarrier
\FloatBarrier
\subsubsection*{TL (III).}
Binary classification between digits ``1'' and ``2'' (source) and ``0'' and ``8'' (target).
\vspace{1mm}
\vspace{1mm}

\begin{table}[H]
\centering
\footnotesize
\setlength{\tabcolsep}{4pt}
\caption{Best accuracies for TL (III) with $n=3$.}
\label{tab:tl3_n3}
\resizebox{\columnwidth}{!}{%
\begin{tabular}{c|c|c|c|c|c}
\hline
 & QCNN-N & QCNN-Z & QCNN-G & CCNN-A & CCNN-B \\
\hline
Large data & 81.06\% (3) & 84.34\% (3) & \textit{89.94\%}(2) & 82.47\% (3) & \textbf{96.98\%} (3) \\
Small data & \textit{89.75\%} (3) & 85.94\%(3) & 85.94\% (2) & 83.29\% (3) & \textbf{93.07\%} (3) \\
\hline
\end{tabular}%
}
\end{table}

\begin{table}[H]
\centering
\footnotesize
\setlength{\tabcolsep}{4pt}
\caption{Best accuracies for TL (III) with $n=4$.}
\label{tab:tl3_n4}
\resizebox{\columnwidth}{!}{%
\begin{tabular}{c|c|c|c|c|c}
\hline
 & QCNN-N & QCNN-Z & QCNN-G & CCNN-A & CCNN-B \\
\hline
Large data & 96.44\% (1) & \textit{97.44\%} (2) & 94.69\% (4) & 85.82\% (4) & \textbf{98.67\%} (3) \\
Small data & \textit{\textbf{97.31\%}} (2) & 96.72\% (2) & 94.94\% (4) & 74.39\% (4) & 93.33\% (4) \\
\hline
\end{tabular}%
}
\end{table}
\FloatBarrier

\subsubsection*{TL (IV).}
Binary classification between digits ``5'' and ``7'' (source) and ``0'' and ``8'' (target).
\vspace{1mm}
\vspace{1mm}

\begin{table}[H]
\centering
\footnotesize
\setlength{\tabcolsep}{4pt}
\caption{Best accuracies for TL (IV) with $n=3$.}
\label{tab:tl4_n3}
\resizebox{\columnwidth}{!}{%
\begin{tabular}{c|c|c|c|c|c}
\hline
 & QCNN-N & QCNN-Z & QCNN-G & CCNN-A & CCNN-B \\
\hline
Large data & \textit{87.87\%} (2) & 87.19\% (3) & 86.12\% (3) & 82.47\% (3) & \textbf{95.39\%} (2) \\
Small data & 87.81\% (3) & 80.88\% (3) & \textit{\textbf{91.75\%}} (3) & 83.29\% (3) & 89.07\% (3) \\
\hline
\end{tabular}%
}
\end{table}

\begin{table}[H]
\centering
\footnotesize
\setlength{\tabcolsep}{4pt}
\caption{Best accuracies for TL (IV) with $n=4$.}
\label{tab:tl4_n4}
\resizebox{\columnwidth}{!}{%
\begin{tabular}{c|c|c|c|c|c}
\hline
 & QCNN-N & QCNN-Z & QCNN-G & CCNN-A & CCNN-B \\
\hline
Large data & 92.88\% (4) & \textit{93.12\%} (4) & 87.50\% (3) & 85.63\% (4) & \textbf{98.11\%} (4) \\
Small data & \textit{92.12\%} (4) & \textit{\textbf{97.31\%}} (4) & 87.69\% (3) & 70.68\% (4) & 91.66\% (3) \\
\hline
\end{tabular}%
}
\end{table}
\FloatBarrier

\FloatBarrier
\subsubsection*{TL (V).}
Binary classification between Fashion-MNIST classes "T-shirt/top" and "Trouser" (source) and ``0'' and ``8'' (target).
\vspace{1mm}
\vspace{1mm}

\begin{table}[H]
\centering
\footnotesize
\setlength{\tabcolsep}{4pt}
\caption{Best accuracies for TL (V) with $n=3$.}
\label{tab:tl5_n3}
\resizebox{\columnwidth}{!}{%
\begin{tabular}{c|c|c|c|c|c}
\hline
 & QCNN-N & QCNN-Z & QCNN-G & CCNN-A & CCNN-B \\
\hline
Large data & \textit{91.06\%} (3) & 88.94\% (3) & 88.38\% (3) & 87.65\% (3) & \textbf{96.93\%} (3) \\
Small data & \textit{86.56\%} (2) & \textit{84.00\%} (3) & 86.25\% (3) & 83.30\% (3) & \textbf{91.11\%} (3) \\
\hline
\end{tabular}%
}
\end{table}

\begin{table}[H]
\centering
\footnotesize
\setlength{\tabcolsep}{4pt}
\caption{Best accuracies for TL (V) with $n=4$.}
\label{tab:tl5_n4}
\resizebox{\columnwidth}{!}{%
\begin{tabular}{c|c|c|c|c|c}
\hline
 & QCNN-N & QCNN-Z & QCNN-G & CCNN-A & CCNN-B \\
\hline
Large data & 94.44\% (4) & \textit{96.15\%} (4) & 93.06\% (4) & 79.34\% (2) & \textbf{99.08\%} (3) \\
Small data & 94.88\% (4) & \textit{\textbf{96.44\%}} (4) & 88.69\% (4) & 76.74\% (0) & 91.81\% (4) \\
\hline
\end{tabular}%
}
\end{table}
\FloatBarrier

\FloatBarrier
\subsubsection*{TL (VI).}
The source task is multi-class classification over digits ``1'' --``7''  and ``9''  in MNIST. The pre-trained model is then transferred to a binary classification task ("0" vs "8") on the target dataset.
\vspace{1mm}
\vspace{1mm}

\begin{table}[H]
\centering
\footnotesize
\setlength{\tabcolsep}{4pt}
\caption{Best accuracies for TL (VI) with $n=3$.}
\label{tab:tl6_n3}
\resizebox{\columnwidth}{!}{%
\begin{tabular}{c|c|c|c|c|c}
\hline
 & QCNN-N & QCNN-Z & QCNN-G & CCNN-A & CCNN-B \\
\hline
Large data & 80.62\% (2) & 89.09\% (3) & \textit{91.06\%} (3) & 72.17\% (2) & \textbf{98.65\%} (3) \\
Small data & \textit{85.88\%} (3) & 77.59\% (3) & 77.19\% (2) & 66.38\% (2) & \textbf{91.06\%} (2) \\
\hline
\end{tabular}%
}
\end{table}

\begin{table}[H]
\centering
\footnotesize
\setlength{\tabcolsep}{4pt}
\caption{Best accuracies for TL (VI) with $n=4$.}
\label{tab:tl6_n4}
\resizebox{\columnwidth}{!}{%
\begin{tabular}{c|c|c|c|c|c}
\hline
 & QCNN-N & QCNN-Z & QCNN-G & CCNN-A & CCNN-B \\
\hline
Large data & \textit{97.19\%} (4) & 95.81\% (3) & 94.19\% (4) & 89.11\% (4) & \textbf{98.63\%} (3) \\
Small data & \textit{\textbf{97.12\%}} (4) & 90.53\% (3) & 92.56\% (4) & 75.08\% (4) & 92.85\% (4) \\
\hline
\end{tabular}%
}
\end{table}
\FloatBarrier

\FloatBarrier
\subsubsection*{TL (VII).}
The source task is multi-classification over all features in Fashion MNIST dataset. The pre-trained model is then transferred to a binary classification task ("0" vs "8") on the target dataset.
\vspace{1mm}
\vspace{1mm}

\begin{table}[H]
\centering
\footnotesize
\setlength{\tabcolsep}{4pt}
\caption{Best accuracies for TL (VII) with $n=3$.}
\label{tab:tl7_n3}
\resizebox{\columnwidth}{!}{%
\begin{tabular}{c|c|c|c|c|c}
\hline
 & QCNN-N & QCNN-Z & QCNN-G & CCNN-A & CCNN-B \\
\hline
Large data & 86.88\% (1) & \textit{89.38\%} (2) & 86.38\% (3) & 87.65\% (3) & \textbf{96.30\%} (3) \\
Small data & 87.44\% (3) & 86.66\% (2) & \textit{88.94\%} (3) & 67.77\% (3) & \textbf{93.58\%} (3) \\
\hline
\end{tabular}%
}
\end{table}

\begin{table}[H]
\centering
\footnotesize
\setlength{\tabcolsep}{4pt}
\caption{Best accuracies for TL (VII) with $n=4$.}
\label{tab:tl7_n4}
\resizebox{\columnwidth}{!}{%
\begin{tabular}{c|c|c|c|c|c}
\hline
 & QCNN-N & QCNN-Z & QCNN-G & CCNN-A & CCNN-B \\
\hline
Large data & \textit{97.00\%} (4) & 98.00\% (3) & 95.44\% (4) & 77.88\% (3) & \textbf{98.78\%} (3) \\
Small data & 88.56\% (4) & 94.62\% (3) & \textit{\textbf{96.56\%}} (4) & 84.20\% (4) & 93.51\% (3) \\
\hline
\end{tabular}%
}
\end{table}
\FloatBarrier


\clearpage
\begin{thebibliography}{99}          
%
\bibitem {survey}Sinno Jialin Pan and Qiang Yang, A Survey on Transfer
Learning, IEEE Trans. Knowl. Data Eng,\textbf{ 22}, 1345 (2009).

\bibitem {s1}Yihao Xue, Rui Yang, Xiaohan Chen, Weibo Liu, Zidong Wang, and
Xiaohui Liu, A Review on transferability estimation in deep transfer learning,
IEEE Trans. Artif. Intell. \textbf{5}, 5894 (2024).

\bibitem {few}Jason Yosinski, Jeff Clune, Yoshua Bengio, and Hod Lipson, How
transferable are features in deep neural networks? In Advances in Neural Information Processing Systems \textbf{4} 3320 (2014).

\bibitem {f1}Haytham M. Fayek Lawrence Cavedon, and Hong Ren Wu, On the
transferability of representations in neural networks between datasets and
tasks, arXiv:1811.12273.

\bibitem {f2}Nima Tajbakhsh, Jae Y. Shin, Suryakanth R. Gurudu, R. Todd Hurst,
Christopher B. Kendall, Michael B. Gotway, and Jianming Liang Convolutional
neural networks for medical image analysis: Full training or fine tuning? IEEE
Trans. Med. Imaging, \textbf{35}, 1299 (2016).

\bibitem {f3}Yixiong Chen, Jingxian Li, Chris Ding, and Li Liu, Rethinking Two
Consensuses of Transferability in Deep Learning, arXiv:2212.00399.

\bibitem {UTL1}Gabriel Michau and Olga Fink, Unsupervised transfer learning
for anomaly detection: application to complementary operating condition
transfer,  Knowl‑Based Syst. \textbf{216}, 106816 (2021).

\bibitem {UTL2}Rakesh Kumar Sanodiya and Leehter Yao, A novel unsupervised
globality-locality preserving projections in transfer learning, Image Vis.
Comput. \textbf{90}, 103802 (2019).

\bibitem {UTL3}Sanodiya R.K., Yao L. Unsupervised transfer learning via
relative distance comparisons, IEEE Access, \textbf{8}, 110290 (2020).

\bibitem {UTL4}Hang Chang, Ju Han, Cheng Zhong, Antoine M. Snijders, and
Jian-Hua Mao, Unsupervised transfer learning via multi-scale convolutional
sparse coding for biomedical applications, IEEE Trans. Pattern Anal. Mach.
Intell. \textbf{40}, 1182 ((2017).

\bibitem {UTL5}Bo Du, Liangpei Zhang, Dacheng Tao, and Dengyi Zhang,
Unsupervised transfer learning for target detection from hyperspectral images,
Neurocomputing, \textbf{120}, 72 (2013).

\bibitem {UTL6}Chenguang Duan, Yuling Jiao, Huazhen Lin, Wensen Ma, and Jerry
Zhijian Yang, Unsupervised transfer learning via adversarial contrastive
training, arXiv:2408.08533.

\bibitem {UTL7}Ruoshi Qin, Feiya Lv, Huawei Ye, and Jinsong Zhao, Unsupervised
transfer learning for fault diagnosis across similar chemical processes,
Process Saf. Environ. Prot., \textbf{190}, 1011(2024).

\bibitem {1}P. W. Shor, Polynomial-time algorithms for prime factorization and
discrete logarithms on a quantum computer, SIAM Rev. \textbf{41} 303 (1999).

\bibitem {2}L. K. Grover, Quantum mechanics helps in searching for a needle in
a haystack. Phys. Rev. Lett. \textbf{79}, 325 (1997).

\bibitem {3}F. Arute\textit{ et al.}, Quantum supremacy using a programmable
superconducting processor, Nature \textbf{574}, 505 (2019).

\bibitem {bp}Martin Larocca, Supanut Thanasilp, Samson Wang, Kunal Sharma,
Jacob Biamonte, Patrick J. Coles, Lukasz Cincio, Jarrod R. McClean, Zo\"{e}
Holmes, and M. Cerezo, A Review of Barren Plateaus in Variational Quantum
Computing, arXiv:2405.0078.

\bibitem {bp1}Jarrod R. McClean, Sergio Boixo, Vadim N. Smelyanskiy, Ryan
Babbush, and Hartmut Neven, Barren plateaus in quantum neural network training
landscapes, Nat. Commun.\textbf{ 9}, 4812 (2018).

\bibitem {bp2}Han Qi, Lei Wang, Hongsheng Zhu, Abdullah Gani, and Changqing
Gong, The barren plateaus of quantum neural networks: review, taxonomy and
trends, Quantum Inf. Process. \textbf{22}, 435, (2023).

\bibitem {qcnn}Iris Cong, Soonwon Choi, and Mikhail D. Lukin, Quantum
convolutional neural networks, Nat. Phys. \textbf{15}, 1273 (2019).

\bibitem {qcnn1}Seunghyeok Oh, Jaeho Choi, Joongheon Kim, A Tutorial on
Quantum Convolutional Neural Networks (QCNN), arXiv:2009.09423.

\bibitem {bp31}Arthur Pesah, M. Cerezo, Samson Wang, Tyler Volkoff, Andrew T.
Sornborger, Patrick J. Coles, Absence of Barren Plateaus in Quantum
Convolutional Neural Networks, Phys. Rev. X \textbf{11}, 041011 (2021).

\bibitem {bp3}Edward Grant, Marcello Benedetti, Shuxiang Cao, Andrew Hallam,
Joshua Lockhart, Vid Stojevic, Andrew G. Green, Simone Severini, Hierarchical
quantum classifiers, npj Quantum Inf. \textbf{4} 65 (2018).

\bibitem {bp4}Arthur Pesah, M. Cerezo, Samson Wang, Tyler Volkoff, Andrew T.
Sornborger, Patrick J. Coles, Absence of barren plateaus in quantum
convolutional neural networks, Phys. Rev. X \textbf{11} 041011 (2021).

\bibitem {good}Tak Hur, Leeseok Kim, and Daniel K. Park. Quantum convolutional
neural network for classical data classification, Quantum Machine
Intelligence, \textbf{4}, 3 (2022).

\bibitem {qcnn10}Andrea Matic, Maureen Monnet, Jeanette Miriam Lorenz,
Balthasar Schachtner, and Thomas Messerer, Quantum-classical convolutional
neural networks in radiological image classification, in Proc. IEEE Int. Conf.
Quantum Comput. Eng. (QCE), 56 (2022).

\bibitem {qcnn11}Fan Fan, Yilei Shi, Tobias Guggemos, and Xiao Xiang Zhu,
Hybrid quantum-classical convolutional neural network model for image
classification, IEEE Trans. Neural Netw. Learn. Syst. \textbf{18} (2023).

\bibitem {qcnn12}Aijuan Wang, Jianglong Hu, Shiyue Zhang, and Lusi Li, Shallow
hybrid quantum-classical convolutional neural network model for image
classification, Quan. Inf. Proc. \textbf{23}, 17 (2024).

\bibitem {qcnn4}Shangshang Shi, Zhimin Wang, Jiaxin Li, Yanan Li, Ruimin
Shang, Guoqiang Zhong, and Yongjian Gu, Quantum convolutional neural networks
for multiclass image classification, Quan. Inf. Proc.\textbf{ 23} 189 (2024).

\bibitem {qcnn6}Kummari Venkatesh, K. Jairam Naik, and Achyut Shankar ,
Quantum convolution neural network for multi-nutrient detection and stress
identification in plant leaves, Multimed. Tools. Appl. \textbf{83} 65663 (2024).

\bibitem {qcnn9}Sanjeev Bhatta and Ji Dang, Multiclass seismic damage detection of
buildings using quantum convolutional neural network, Comput.-Aided Civ. Infrastruct. Eng. \textbf{39}, 406 (2024).

\bibitem {ggood}Li-An Lo, Li-Yi Hsu, and En-Jui Kuo, Unsupervised Feature
Extraction and Reconstruction Using Parameterized Quantum Circuits, arXiv:2502.07667.

\bibitem {cq}Juhyeon Kim, Joonsuk Huh, and Daniel K. Park,
Classical-to-quantum convolutional neural network transfer learning,
Neurocomputing \textbf{555} 126643 (2023).

\bibitem {qo}Andrea Mari, Thomas R. Bromley, Josh Izaac, Maria Schuld, and
Nathan Killoran, Transfer learning in hybrid classical quantum neural
networks, Quantum \textbf{4}, 340 (2020).

\bibitem {cq1}Emmanuel Ovalle-Magallanes, Juan Gabriel Avina-Cervantes, Ivan
Cruz-Aceves, and Jose Ruiz-Pinales, Hybrid classical-quantum Convolutional
Neural Network for stenosis detection in X-ray coronary angiography, Expert
Syst. Appl. \textbf{189}, 116112 (2022).

\bibitem {cq2}Vanda Azevedo, Carla Silva, In\^{e}s Dutra, Quantum transfer
learning for breast cancer detection, Quantum Mach. Intell.\textbf{ 4}, 5 (2022).

\bibitem {cq3}Angelina Gokhale, Mandaar B. Pande,\ and Dhanya Pramod,
Implementation of a quantum transfer learning approach to image splicing
detection, Int. J. Quantum Inform. \textbf{18}, 2050024 (2020).

\bibitem {cq4}Amena Khatun and Muhammad Usman, Quantum transfer learning with
adversarial robustness for classification of high-resolution image datasets,
Adv Quantum Technol. \textbf{8}, 2400268 (2025).

\bibitem {cq5}Jun Qi and Javier Tejedor, Classical-to-quantum transfer
learning for spoken command recognition based on quantum neural networks, In
Proc. IEEE Int. Conf. Acoust. Speech Signal Process. (ICASSP) 8627 (2022).

\bibitem {cq6}Shouwei Hu, Xi Li, Banyao Ruan, Zhihao Liu, An
Amplitude-Encoding-Based Classical-Quantum Transfer Learning framework:
Outperforming Classical Methods in Image Recognition, arXiv: 2502.20184.

\bibitem {qc1}Yumin Dong, Xuanxuan Che, Yanying Fu, Hengrui Liu, Yang Zhang
and Yong T, Classification of knee osteoarthritis based on
quantum-to-classical transfer learning, Front. Phys. \textbf{11}, 1212373 (2023).

\bibitem {qc}Anthony M. Smaldone and Victor S. Batista, Quantum to classical
neural Network transfer learning applied to drug toxicity prediction, J. Chem.
Theory Comput.\textbf{ 20}, 4901 (2024).

\bibitem {a2}Ian MacCormack, Conor Delaney, Alexey Galda, Nidhi Aggarwal,
Prineha Narang, Branching Quantum Convolutional Neural Networks, Phys. Rev.
Res. \textbf{4}, 013117 (2022).

\bibitem {a3}Changwon Lee, Israel F. Araujo, Dongha Kim, Junghan Lee, Siheon
Park, Ju-Young Ryu, Daniel K. Park, Optimizing quantum convolutional neural
network architectures for arbitrary data dimension, Front. Phys. \textbf{13},
1529188 (2025).

\bibitem {a1}Farrokh Vatan and Colin Williams. Optimal quantum circuits for
general two-qubit gates. Phys. Rev. A, \textbf{69}, 032315 (2004).

\bibitem {tol}R. S. Woodworth and E. L. Thorndike, The influence of
improvement in one mental function upon the efficiency of other functions.
(I), Psychol. Rev., \textbf{8}(3) 247 (1901).

\bibitem{quantinuum_h1}
Quantinuum H2-1.  \url{https://www.quantinuum.com/}, 
Experiment dates: December, 02-03, 2025.

\bibitem{SVM}
C. Cortes and V. Vapnik,
Support-vector networks,
Machine Learning 20, 273--297 (1995).


\bibitem{KTA}
N. Cristianini, J. Shawe-Taylor, A. Elisseeff, and J. Kandola,
On kernel-target alignment,
Advances in Neural Information Processing Systems
\textbf{14}, 367--373 (2001).


\bibitem{KTA_review}
T. Wang, D. Zhao, and S. Tian,
An overview of kernel alignment and its applications,
Artif. Intell. Rev. \textbf{43}, 179--192 (2015).



\end{thebibliography}
\end{document}